# In Pursuit of Unification of Conceptual Models: Sets as Machines

Sabah Al-Fedaghi*
*Computer Engineering Department*
*Kuwait University*
*Kuwait*
salfedaghi@yahoo.com, sabah.alfedaghi@ku.edu.kw

*Abstract* - Conceptual models as representations of real-world systems are based on diverse techniques in various disciplines but lack a framework that provides multidisciplinary ontological *understanding* of real-world phenomena. Concurrently, systems' complexity has intensified, leading to a rise in developing models using different formalisms and diverse representations even within a single domain. Conceptual models have become larger; languages tend to acquire more features, and it is not unusual to use different modeling languages for different components. This diversity has caused problems with consistency between models and incompatibly with designed systems. Two main solutions have been adopted over the last few years: (1) A currently dominant technology-based solution tries "to harmonize or unify" models, e.g., unifies EER and UML. This solution would solidify modeling achievements, reaping benefits from huge investments over the last thirty years. (2) A less prevalent solution is to pursue deeper roots that reveal unifying modeling principles and apparatuses. An example of the second method is a "category theory"-based approach that utilizes the strengths of the graph and set theory, along with other topological tools. This manuscript is a sequel in a research venture that belongs to the second approach and uses a model called thinging machines (TMs) founded on Stoic ontology and Lupascian logic. TM modeling contests the thesis that there is no universal approach that covers all aspects of an application, and the paper demonstrates that pursuing such universality is anything but a dead-end method. This paper continues in this direction, with emphasis on TM foundation (e.g., existence and subsistence of things) and exemplifies this pursuit by proposing an alternative representation of set theory.

## I. Introduction

Conceptual modeling is a crucial aspect of development for software, systems, and knowledge engineering. According to [1], the software development process is a kind of problem-solving process that requires understanding of all problem components, relations, rules, constraints, etc. Such an understanding "is a hard and time-consuming process, which requires specialized tools for being performed. These tools, which allow the software engineer to understand the problem to solve, are known as conceptual models" [1].

------------------------------------------
*Retired June 2021, seconded fall semester 2021/2022

Over the past four decades, the field of conceptual modeling has continued to evolve to be applied to important problems in many disciplines [2]. It has shifted from being a software engineering technique to being "a standalone discipline that has a value proposition for any domain […] where complexity must be managed through abstraction and structuring" [3].

In conceptual modeling, a variety of modeling languages were used first, giving rise to overlap of common concepts and notions across the various modeling languages. This revealed the need for *a unification* of the various languages, notions, or models. Pursuing such a course resolved into a standardization effort first called the "Unified Method" and later the "Unified Modeling Language" (UML) [4]. UML arose from the unification of several object-oriented design methods. According to [4], "during this crucial period of unification, it became clear that defining such a standard would not be an easy task. A research community emerged that became interested in studying models and the UML as a core research subject area of its own." In their work, [5] use multiple models in their systems analysis and design practices; however, there is no comprehensive theory yet to explain how practitioners would work with these models [5].

The concept was broadened to the general issues of modeling languages as a core subject of study that is not specifically tied to the UML [4]. Some experts call for "more and deeper studies of [UML's] longer-term use in the field" [6].

### A. Problem: Fragmentation

Extending the general issues related to conceptual modeling, this type of modeling has fragmented methods in *various disciplines,* and there is no common unifying framework that shows how concepts come together to represent the target system. Additionally, systems' complexity has increased, leading to a rise in developing models with different intents that are written using various formalisms that give diverse system representations [7]. With increasingly complex system development and the need for systems integration, conceptual models have become larger; their languages tend to acquire more features, and it is not unusual to use different modeling languages for different components (ER + UML) [7].

In general, models are used to support a number of purposes such as construction of systems, communication, analysis, documentation, evolution, realization, and construction [8]. According to [8], we might overburden a model to satisfy all these purposes. Instead, we may use a number of models for each of its purposes and then bind these models to each other. Diversity can lead to problems with consistency between models and incompatibility with the designed system. This is especially true of complex systems that can interact with humans [7].

*B. In Pursuit of Unification*

Accordingly, in this paper, the goal is locating a possible unifying representation that is expressive enough to handle diverse modeling notions. Currently, two undertakings are directed for such a purpose. The first is technological-based efforts "to harmonize or *unify* them" [9], described by [3] as techniques "subordinated" to a certain discipline (e.g., software engineering). This approach is exemplified by presently available modeling languages, e.g., EER and UML-class diagrams [10], with the aim of developing full transformation among various diagrammatic representations.

The second approach involves an attempt to develop a unified modeling methodology from scratch to provide a conceptual framework that may or may not apply to the current multiplicity of models. An example of such a quest is Reference [7]'s work, which proposed a *category theory* that "offers a point of view that allows us to use both the strengths of the graph and set theories, along with other topological tools […] to provide a mathematical framework for synchronization methodologies" (italics added). Category theory is an endeavor to formalize various mathematical structures and their relationships. Reference [7] used graphs and set theories to provide a framework for a unified representation of conceptual models. Category theory seems to be a potential tool for various unification efforts. Reference [11] proposed using category theory to develop multi-model data representation for transformations between models.

This manuscript is a sequel in a research venture that belongs to the second approach to conceptual modeling unification, aimed at exemplifying a modeling language that promises an initial unifying thesis to disprove the common claim that one model that can describe every application does not exist. Of course, such a language has its limitations, but it describes all that can be described.

In a way, our approach contests the thesis [8] that "there is no universal approach and no universal language that covers all aspects of an application that have a well-founded semantics for all constructions, which reflect any relevant facet in applications, and that support engineering." Admittedly, our proposed modeling methodology may not completely satisfy Thalheim [8]'s requirements; nevertheless, it shows that pursuing universal modeling approach is anything but a dead end method.

We use a model called thinging machines (TMs) founded on Stoic ontology and Lupascian logic [12]. TM modeling contests the thesis that there is no universal approach and no universal language that covers all aspects of an application. TMs have been applied in many applications (networks, hardware, systems, story-telling, communication, railcar systems, robotics, business, etc.) using a single diagrammatic language based the notion of a TM. Note that UML has been used to model such applications, but UML uses 14 diagrammatic languages (e.g., activity, sequence, state, class, […] diagrams). This paper continues in this direction with emphasis on representations in the context of set theory.

*C. The Paper's Structure*

The next section includes a brief review of TM modeling with some new clarifications. Section three contains an example of TM modeling of the process of car hire. Section four includes theoretical issues in TM modeling, including potentiality, actuality, existence, and subsistence. Section five is focused on some aspects of set theory representation.

## II. TM MODEL

The TM model is based on the supposition that "there is a ready-made world" [13] that reflects some "fundamental," "joint-carving," or "structural" concepts. A complete description of reality using these concepts gives reality's fundamental structure. According to [8], concepts specify "what things are there and what properties things have." *Structure* reveals where the joints of the world can be carved, because "structure is the right and proper way to find these joints, and go about this carving" ([14] referencing the philosopher Theodore Sider).

The TM model's basic "carving at the joints" of reality produces what is called a *thimac* (**thi**ng/**mac**hine). A thimac has a dual nature of being as a *thing* and simultaneously as a *machine*. The goal of such duality is an attempt to create a unifying notion that represents "entity-ness" and "process-ness." A thimac is a *machine* when it acts (subject) on other thimacs, and it is a *thing* when it is the object of actions by other thimacs. The thimac (and subthimacs) also has a dual mode of reality (Stoic idea): *subsistence* static (timeless) reality and *existence* event-based reality. In the context of making models, and demonstrated in systems and software engineering, TMs emphasize that entities and processes are viewed as thimacs in what may be called the *TM universe*. Physical particulars (e.g., a cat curled up on a sofa) and sets (will be demonstrated in this paper for logic-based sets) can be represented uniformly as thimacs.

## A. Things and Machine

As will be argued latter, grounded on the Stoic ideas, the world of existence spurts out of a world of subsistence. Subsistence is the totality of timeless thimacs. In this case, the subsistence of a thimac (its static level representation) "involves once for all everything that will ever happen to" that thimac (the quoted expression is borrowed from a description of Leibniz's work [15]).

The thimac *machine* consists of five actions: *create*, *process*, *release*, *transfer* and *receive*. (See Fig. 1). The thimac *thing* is whatever created, processed, released, transferred, and received. A thimac as a machine creates, processes, releases, transfers, and receives.

TMs' actions are described as follows.
1) *Arrive:* A thing arrives to a machine.
2) *Accept:* A thing enters the machine. For simplification, we assume that arriving things are *accepted* (see Fig. 1); therefore, we can combine the *arrive* and *accept* stages into the *receive* stage.
3) *Release:* A thing is ready for transfer outside the machine.
4) *Process:* A thing is changed, handled, and examined, but no new thing results.
5) *Transfer:* A thing is input into or output from a machine.
6) *Create:* A new thing is manifested in a machine.

Additionally, the TM model includes a *triggering* mechanism (denoted by a dashed arrow in this article's figures), which initiates a (nonsequential) flow. Moreover, each action may have its own storage (denoted by a cylinder in the TM diagram). For simplicity, we may omit *create* from some diagrams because the box representing the thimac implies its being-ness (in the model). Fig. 2 shows the set of TM modeling notations.

**Example**: Fig. 3 shows an illustrative representation of a car as a *thing* and as a *machine*.

## B. Two-level Modeling

TM modeling involves a representation with vertically dynamic depiction over a *timelessness* (static) picture (see the coin illustration in Fig. 4). In a TM, reality has two modes: subsistence and existence.

The static model is built from subsisting *regions* (subdiagram of the static description) with a logical order imposed by *potential* flows and triggers. The static model comprises fixed parts, and it simply *subsists*, e.g., "the flow of *traffic* depends on cars [and flow] without being anything but the cars" [16]. Traffic is not itself a solid body, but it is nonetheless *real* because it depends on cars and roads for its *subsistence* in reality; this subsistence is captured in the static region. Cars are entities that exist as physical things. Traffic is a process that subsists as a *region* of the *existing* traffic in the dynamic level. If there are no cars, traffic still subsists as a *potential* thing.

According to the Stoic, subsistence is as real as existence. Heidegger's "ready-to-hand" hammer refers to an *existing* hammer and *existing* hammering process. It is possible that the hammer is too heavy; hence, it exists, but hammering *subsists* as potentiality. The subsistence of hammering is as real as its existence as a real process. It "is there," distinct from the hammer, analogous to the Shakespeare's drama of "a pound of flesh." Try to hit a nail with a hammer without hammering. The hammer itself is in subsistence if the head flies off the hammer (present-at-hand).

## C. Regions

A *region* of thimacs forms a *static* structure (blueprint) in the world. The region is the TM *space*-equivalent of a mesh (net) of thimacs. TM space comprises multi-net multilevel thimacs.

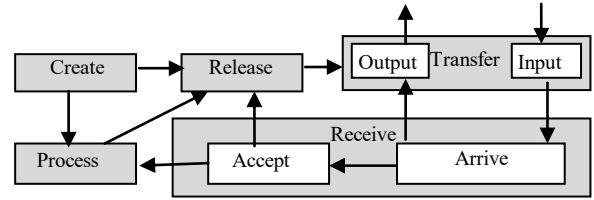

Fig. 1 Thinging machine.

|  | | Function |
|---|---|---|
| Name | Thimac, subthimac, event and region | Boundary context |
| Create, process, release, transfer, or receive | Action | Operation |
| 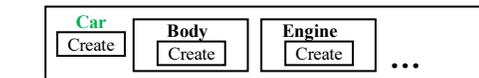 | Flow | Relate/Connect |
| - - - ▶ | Triggering | New flow |

Fig. 2 TM modeling notations.

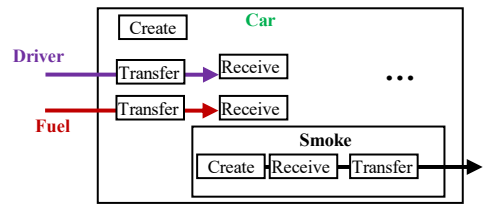

(a) A car as a thing

(b) A car as a machine.

Fig. 3 A car as a thing and a machine.

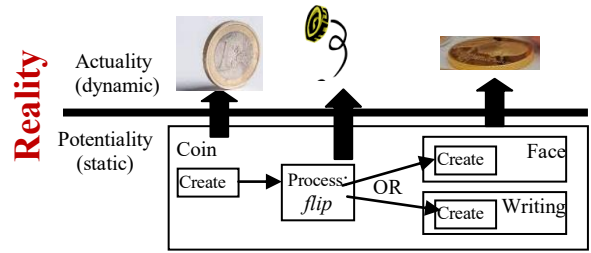

Fig. 4 Potentiality and actuality.

There is no space in this very thick jungle of thimacs. A region may combine with time to form a *dynamic* event. To describe an event precisely, we need the region and time—the combination itself is a thimac.

*D. Existence*

In a TM, a process is an event or a mesh of events, and existence is the flow of events. Thus, existence is a huge process. TM treats activities, objects, and states in a uniform way as events. Objects are nothing more than long events.

The Stoic ontology has two levels of specification: (1) a subsistence static model in which things and actions subsist and (2) an existence dynamic model in which things and actions exist in time. From the Stoic ontological point of view, while a thing existing has a clear denotation, subsistence indicates the thing is "being there," but it is inactive. Thus, *a red apple* refers to a subsisting apple thimac with a redness subthimac, while *red apple on the table now* refers to an event (at a certain time, now) with the region *a red apple on the table*. *Dragons do not exist (now)* refers to a dragon that is not instantiated in existence now. Furthermore, *negative* existence refers to a thimac being in subsistence. Using Lupascian logic to represent negativity is a topic discussed in previous papers.

### III. EXAMPLE OF TM MODELLING

Reference [17] described UML as a *semantically and syntactically rich* visual modeling language for the architecture, design, and implementation of complex software systems, both structurally and behaviorally. A small set of diagrams can be used effectively to model business processes. By having the business analyst and system developers use the same modeling concepts, the risk of costly errors related to different understanding of methodology concepts is significantly mitigated.

According to [17], activity diagrams are easier than other UML diagrams for analysts and stakeholders to fully comprehend. They are the most suitable diagram for business process modeling because they neatly illustrate the flow of a process from activity to activity.

In spite of the claimed *semantical and syntactical richness*, such a simple notion as *activity* does not have a precise definition in UML literature [18]. According to the philosopher Bertrand Russel [15], "Activity is, as a rule, a cover for confused thinking; it is one of those notions which, by appealing to psychological imagination, appear to make things clear, when in reality they merely give an analogy to something familiar."

In a TM, the activity notion is built upon five *generic* (have no sub-actions) actions: create, process, release, transfer, and receive. These are static actions that become (generic) events when they interweave with time.

Reference [17] gave an example of an activity diagram for the process of "car hire." The basic components in Gallia [17]'s example include also the notion of an *event*. The (outside) event is described as a "*process* [that] creates an event but the outcome of the event is outside the scope of the activity diagram" (italics added).

*A. Static Model*

In a TM, an activity as a process is a thimac (with its static and dynamic forms), and an event is a region injected with time. To illustrate these terms, Fig. 5 shows the static model of the *Car Hire* example. In the figure,

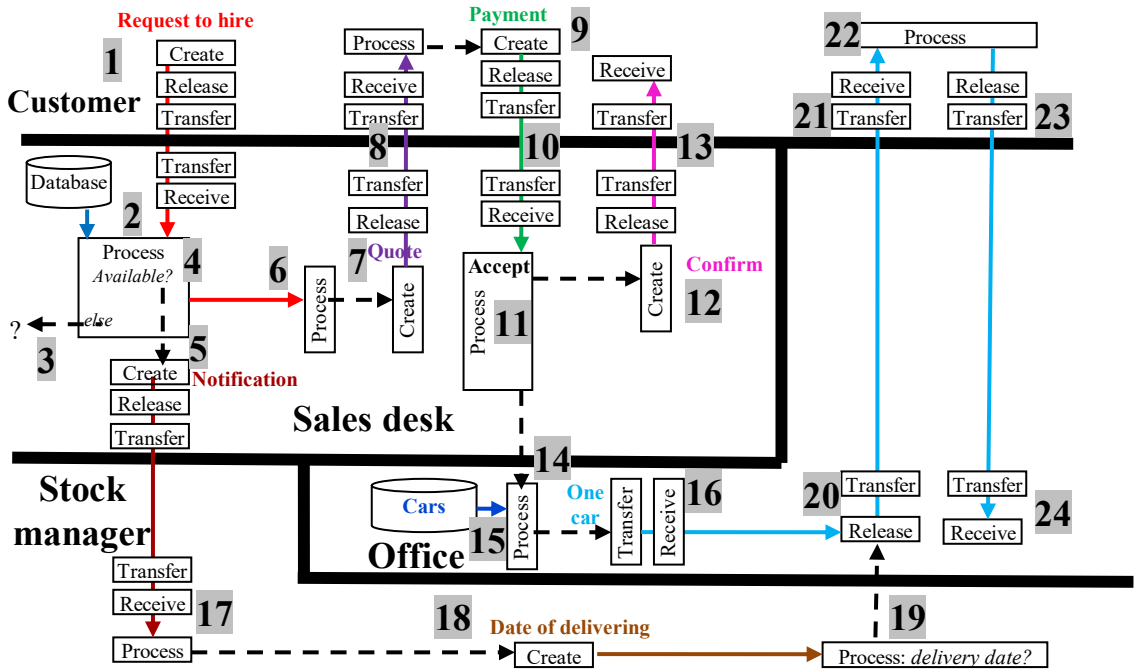

Fig. 5 The static model of the *car hire* system.

- The customer creates a request to hire a car (grey number 1). The request flows to the sales desk, where it is processed (2) by comparing the request with the database of available cars. Accordingly,
  (a) If the car is not available to hire, Gallia [17]'s example does not indicate what to do (3).
  (b) If the car is available (4), then,
    - Trigger a notification that is sent to the stock manager (5).
    - The customer's request is further processed (6) to create the quote for service (7) that flows to the customer. The customer processes the quote (8) and creates a payment (9) that flows to sales desk (10). The payment is processed (11), and if it is accepted, a confirmation (12) is sent to the customer (13). Additionally, the office is triggered to prepare the car (14).
- In the office, the stock of cars is processed (15) to select the appropriate car (16).
    - In parallel, the notification that was sent to the stock manager previously (5) is processed (17). Accordingly, the stock manager sets up the date to deliver the car (18).
    - Accordingly, when the date of delivery arrives, the stock manager triggers (19) delivering the car to the customer by the office (20 and 21).
- The customer uses the car (22) then returns it (23 and 24).

B. *Dynamic Model*

Fig. 5 illustrates the process of describing the organized structure of reality. As mentioned in the famous Platonic metaphor, the world comes to us predivided, and our best method of identifying distinct kinds of things is to carve nature at its joints. This stable view of reality needs to be supplemented with dynamism that "co-exists" with staticity simultaneously to form event thimacs. The implication is that reality is the source of staticity and dynamism. Fig. 6 shows the dynamic model of the car hire system.

In TM modelling, *existence* (a mode of reality beside subsistence) is *being in time* as an event. Consider the *region* subdiagram of Fig. 5 that is shown in Fig. 7. When such a region injected with time, it becomes the event—e.g., *the customer request is sent to the sales desk* (in relative time with respect to other events). For simplification's sake, events are represented by their regions in Fig. 6.

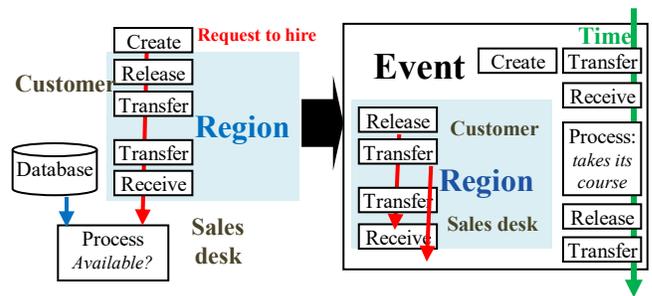

Fig. 7 Static region becomes an event (*the customer request is sent to the sales desk*) when it is injected with time.

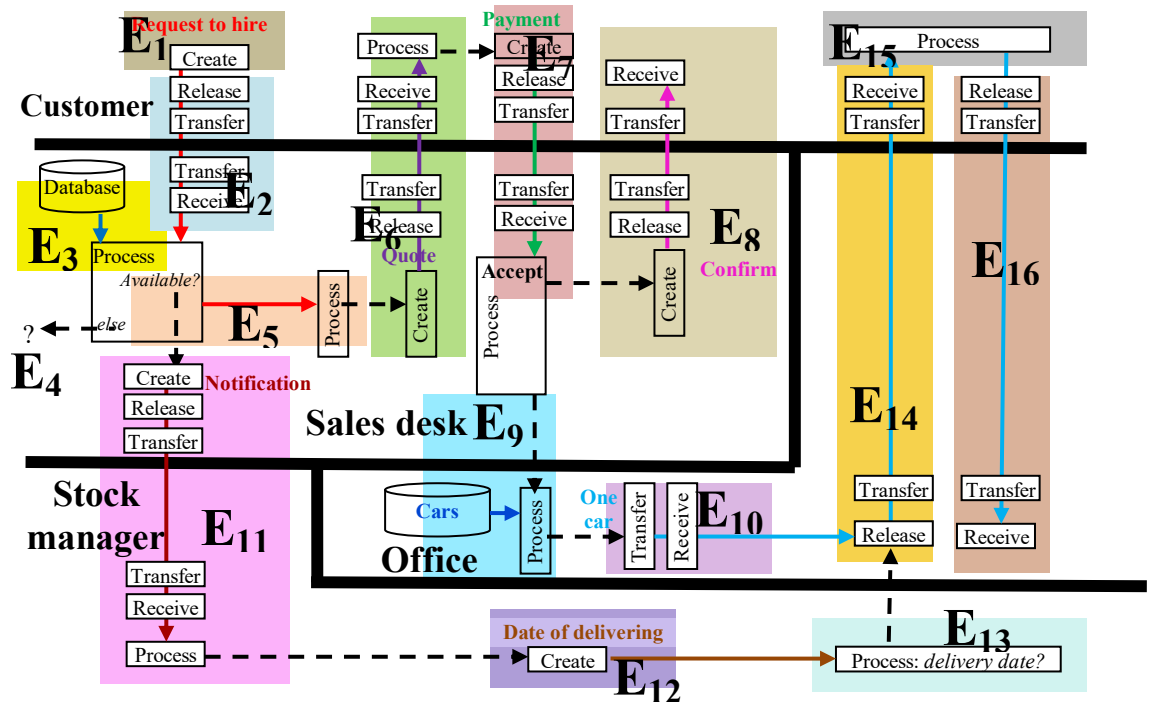

Fig. 6 The dynamic model of the care hire system.

This description also brings to mind Plato's *form* that is generalized in TMs as nets of (static) thimacs, i.e., a subdiagram of the static model. Fig. 8 shows the chronology of events in the car hire example.

IV. THEORETICAL ISSUES IN TM ONTOLOGY

The task of pursuing a unified conceptual model requires involvement in some ontological issues. Ontologies are used to develop new modeling languages, proposing patterns and anti-patterns and improving (semantic) interoperability [19]. This necessitates the utilization of ontological issues from diverse areas (e.g., philosophy and metaphysics, formal ontology, cognitive science, and logics) to develop engineering artifacts for the theory and practice of conceptual modeling [20]. However, in this paper, one may interpret "classical" as focused on whatever is understood to accommodate TM modeling, as historians generally find appropriate.

*A. TM Potentiality and Actuality*

Ontology is an essential topic in the scope of an important area of current computer science and Semantic Web [21]. After the phase of developing a mere modeling language and notations, there comes the issue of providing a framework of ontologically sound models and a more refined depiction of the target systems (what models represent). This is a natural step in studying conceptual modeling, as can be seen from the active research to build ontology for UML (e.g., OntoUML). After all, it is claimed that, "firmly entrenched in many information systems circles […] ontology as 'the specification of a conceptualization' means 'conceptual model'" [22].

Exploring ontological matters leads, first, to old Greece. Aristotle maintains that the world is one of entities and properties. Substance is a class of beings that are actualized as self-standing and independent entities [23] in addition to "accidental" entities such as quality and quantity. Thus, *Fido the dog* is a primary substance—an individual—but *dog* or *doghood* is secondary substance [23]. *Fido* is an object, and *dog* can be predicated of Fido. "Substance" reflects durability or even permanence. *Events* are never substantial in this sense, because they are fleeting [23]. Aristotle postulated two types of thing. The first is measured by time (e.g., movement, processes). The second is material objects in time [24].

The philosophical issue here involves the nature of so-called *accidents* (e.g., motion). The classical view is that an accident was something *in* a body, but nothing without a body, and it cannot subsist of and by itself. The other view considers accidents to be "between body and no body." Accidents are supposed to exist, but to depend upon bodies for their existence [25]. Classically (e.g., Aristotle,

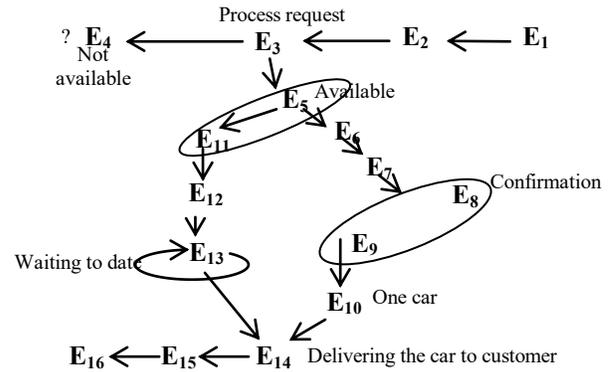

Fig. 8 The behavior model (the ovals indicate parallel processes).

Aquinas), it is said that every physical object is a compound of matter and form. On the other hand, other philosophers claim that form can exist without matter and prime matter can exist without form [26]. According to several references (e.g., [27] and [28]), Aristotle considered that potentiality and actuality are **two kinds of reality and that actualities give origin to potentiality, which gives origin to actualities**. According to [21], Aristotle's actualities are the origin of potentialities, which can generate new actualities.

The purpose of such a discussion is that TMs' two-level model of potentiality and actuality as two aspects of reality is not a new idea. To emphasize the TM's basic ontological claim of two modes of reality, Fig. 9 exemplifies it in terms of *Dog* and *Fido* as discussed previously.

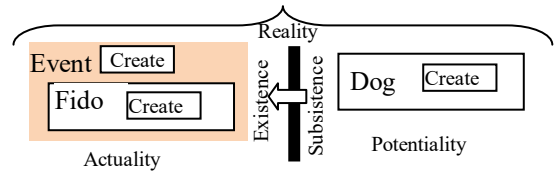

Fig. 9 Dog (Fido) TM model.

According to [21], "The concepts of actuality and potentiality, and of the movement from the latter to the former, have been discussed since Aristotle, but now can be seen as common to both quantum and macroscopic levels of reality. Reference [27] states that the transition from states of potentiality to their actualization is the "basic mechanism of our reality." "State" here refers to a state of affairs. All classical notions such as states, states of affairs (situations), and processes can be represented as TM events.

The nature of potential thimacs and regulating the transition from potential to actual entities and processes are issues in quantum theory. According to [27], consider a particle (e.g., an electron) impinging on a screen. According to quantum mechanics, we cannot know where it will hit, but we can always assign probabilities to the electron's potential to hit at various locations. The entity in consideration is a potential and not an actual entity. It must be considered real and ontologically significant, but not actual. At some time, the electron impinges into some point of the screen, and because it hits the screen, we no longer have a matter of probability.

*B. TM Existence*

In TM modelling, ***Being*** encompasses all thimacs in the two-level representation: existing, subsisting, and those thimacs that cannot materialize in existence (e.g., square circle). A conceptual framework of kinds of things in reality is established from regions said to *subsist* and events said to *exist*. Because all subsisting things are in the static world and all existing things are the regions' counterparts on the dynamic level, regions can *exist* only in the interior of events. For example, the traffic static region only exists in cars and roads, assuming, for simplicity's sake, that cars and roads are the only components of traffic. In another words, the existence of traffic is dependent on the existence of cars and roads.

Reference [29] gives the example of a set of soccer players: "As long as it is just a set of soccer players no *emergence* takes place." This is a static description of subsisting, as shown in Fig. 10 (a). At some point in history, the thimac of a soccer team in its current form arose for the first time in England in the middle of the 19th century. Thus, this thimac has been added to the universe catalog as a subsisting thing that may exist anywhere in the world (potentiality). Subsistence is a kind of registrar of world items as the universe evolves, giving birth to new things. It emerges as a potentiality from the addition of new things to actuality. Reference [29] continues, "Suppose, however, that the set of players starts to practice with the aim of forming a soccer team." In a TM, this is represented by the events shown in Fig. 10 (b). The co-adaptation that takes place between the players during their training and matches results in the emergence of a soccer team, a new structure formed out of the set of individual soccer players.

*C. Origin of Potentiality is Actuality*

According to [30], for a thing to become actual, there must actually exist some endowment of active power or potency, making this possible thing actual. Reality that is produced by and subject to change is a mixture of potential and actual. It is from our *experience of actuality and change that we derive our notion of potential being* as distinct from that of actuality. It is from our experience of what actually exists that we are able to determine what can, and what cannot exist. Static subsistence is discovered in actuality [30]. The two alternating positive states of the same being—in the active existing that produces it and in the pre-existing actual thing—are *real*, in distinction from the mere *logical or objective possibility* of such a being [30].

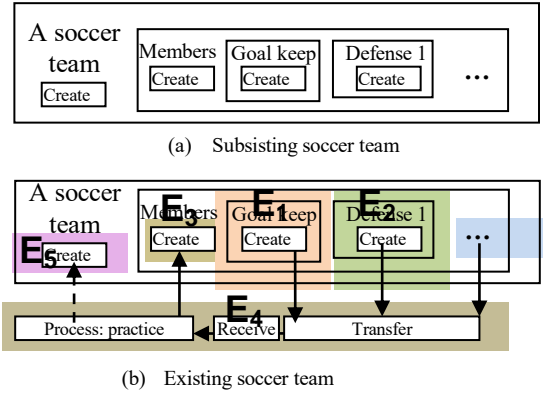

Fig. 10 Creating an existing soccer team.

This discussion implies that potentiality "emerges" from actuality when the latter happens the first time. If an event happens, then its region (structure) becomes part of potentialities. Suppose that the universe's existence started with an event that contained a single, hot, dense point; therefore, the static universe was etched as a static, hot, and dense point as a region of that event. Then, events created hydrogen, helium, and lithium to form heavier elements for the first time. Accordingly, the static description is supplemented with regions of these events. In modern times, roads, cars, and traffic appeared, and their regions became part of the static catalogue of the world to exist (based on their potentialities) time after time.

## V. Set theory

In continuation of applying TMs to present new views of *basic notions* in various types of applications, this section is a preliminary presentation of TMs in set theory. The goal is not to propose a new contribution to set theory; rather, the objective is to explore features of TM modeling. Accordingly, we avoid difficult issues such as infinity and emptiness. In this context, we provide an alternative *representation* of a set as a *thimac* with three *subthimacs, member, extension and transformation* between them.

The notion of sets can be based on logical propositions that represent the set of worlds in which propositions function as the bearers of truth values. A proposition can be associated with a set of "entities" that satisfy the proposition predicate. Thus, for a set S and an object s, we say that S contains s if the following is satisfied: $s_x \in S \longleftrightarrow p(s_x)$ is true.

Because the topic is extensive, we focus on a few aspects of sets in computer science. Future research will extend the treatment to mathematical sets.

*A. Set Theory in Computer Science*

Set theory has been used to represent things related to computer sciences, such as algorithms and designing, data structures and implementations of set operations, database theory, formal language theory, and programming language semantics. Set theory seems to be one of the foundations of computer science and software engineering.

Notions in conceptual modeling are closely related to set theory. For example, a UML class defines a set with members as objects or instances of that class. Based on set theory, [31] presented a formal syntax and semantics for OCL (Object Constraint Language) to allow formally specified constraints on a UML. Reference [32] introduced the B development process of the life cycle of software development. The B development process is utilized to prove that the final code implements its formal specification. B notations are based on set theory and generalized substitutions [32]. Reference [33] introduced an approach for the specification and matching of structural patterns in conceptual models. To build sets representing structural model patterns, [33] defined operations based on set theory, which can be applied to arbitrary sets of model elements and relationships. Reference [34] mapped logical concepts of set theory with UML. Reference [35] formalized "use case" diagrams as "one of the most used diagrams among UML practitioners" based on set theory by logic and quantification. Set theory is utilized for composing and decomposing constructed models (UML, OCL), specifying transformation operations between diagram types and promoting understanding of the system under design [36][37]. Reference [38] applied set operations to merge, slice, and check UML models. A set operation as binary operations $D \times D \rightarrow D$ produces one UML diagram out of two input diagrams for basic set operations: union, intersection, and difference.

### B. Set as a Thimac

As shown in Fig. 11, a set can be conceptualized as a thimac with three subthimacs.
- Member (singularity): one thing
- Extension (multiplicity): a pile of things, disregarding any order or repetition of the things that may be contained within it
- A transformation between extension and member

A member can be receive in the set (number 1), processed (2), and, if qualified, go to transformation (3). Additionally, the extension (4) is sent to the transformation. There, the member and extension are processed (5) to produce a new extension (6).

Similarly, an extension (7) can be sent to the transformation to be processed to select a single member/chalet (8) that is sent to a member (9). The whole extension can be exported (10).

## VI. SAMPLE APPLICATION

In this section, we develop a general system for the chalet market, with minor modifications to Nasef et al. [39]'s problem.

Reference [39] presented an example of application of set theory in a decision-making problem for real estate marketing. The example involves a chalet-selling system in which chalets are described in terms of the following parameters {expensive; beautiful; wooden; cheap; in green surroundings; modern; in good repair; in bad repair}. Each buyer is interested in buying a chalet on the basis of their choice parameters. Out of available chalets, the customer is to select one that meets all of their parameters. Reference [39] used binary tables to develop a method to select a chalet for a customer and applied the method to customers' specific requests. Clearly, the so-called permanents form seven sets. The customer requests an intersection set of some of the seven sets.

### A. Static Model

Fig. 12 shows the TM representation of the corresponding static system. The system involves two main processes as follows.

*Building the seven parameter sets*: This includes adding chalets to various sets according to the seven parameters. For simplification, in Fig. 12, we will factor out the 'Member: Singularity – One' for all sets. An input of the chalet with its description is processed (numbers 1 and 2).

According to the parameters of the chalet, the chalet is added to the relevant set. For example, if the chalet is expensive (3), it goes to the expensive chalet set (4). The set is modeled as a thimac. It involves the extension (ext) (5), a single item (6), and a process that inserts the given item in the set (7). For simplicity's sake, we do not include the item or global process in boxes as subthimacs, as we do for the set.

Accordingly, the input expensive chalet and the current set of expensive chalet are processed (7) to inset the input chalet description to create a new set (5).

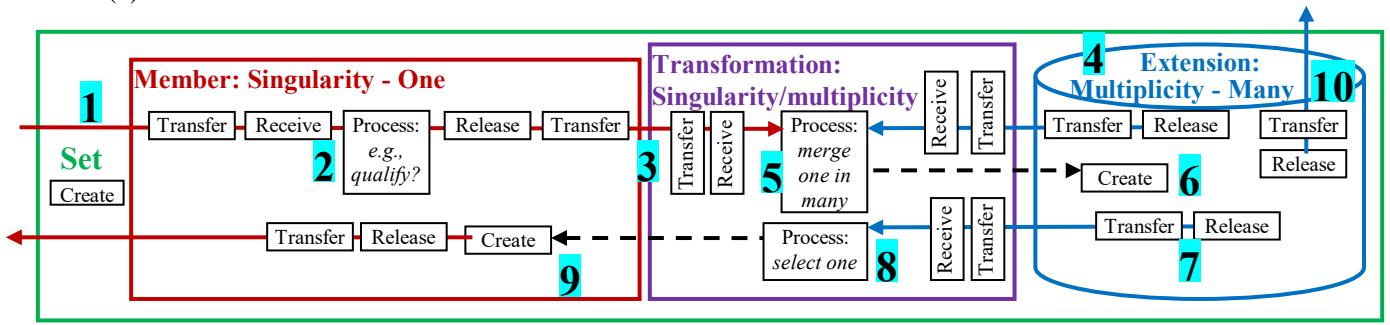

Fig. 11 A set as a machine.

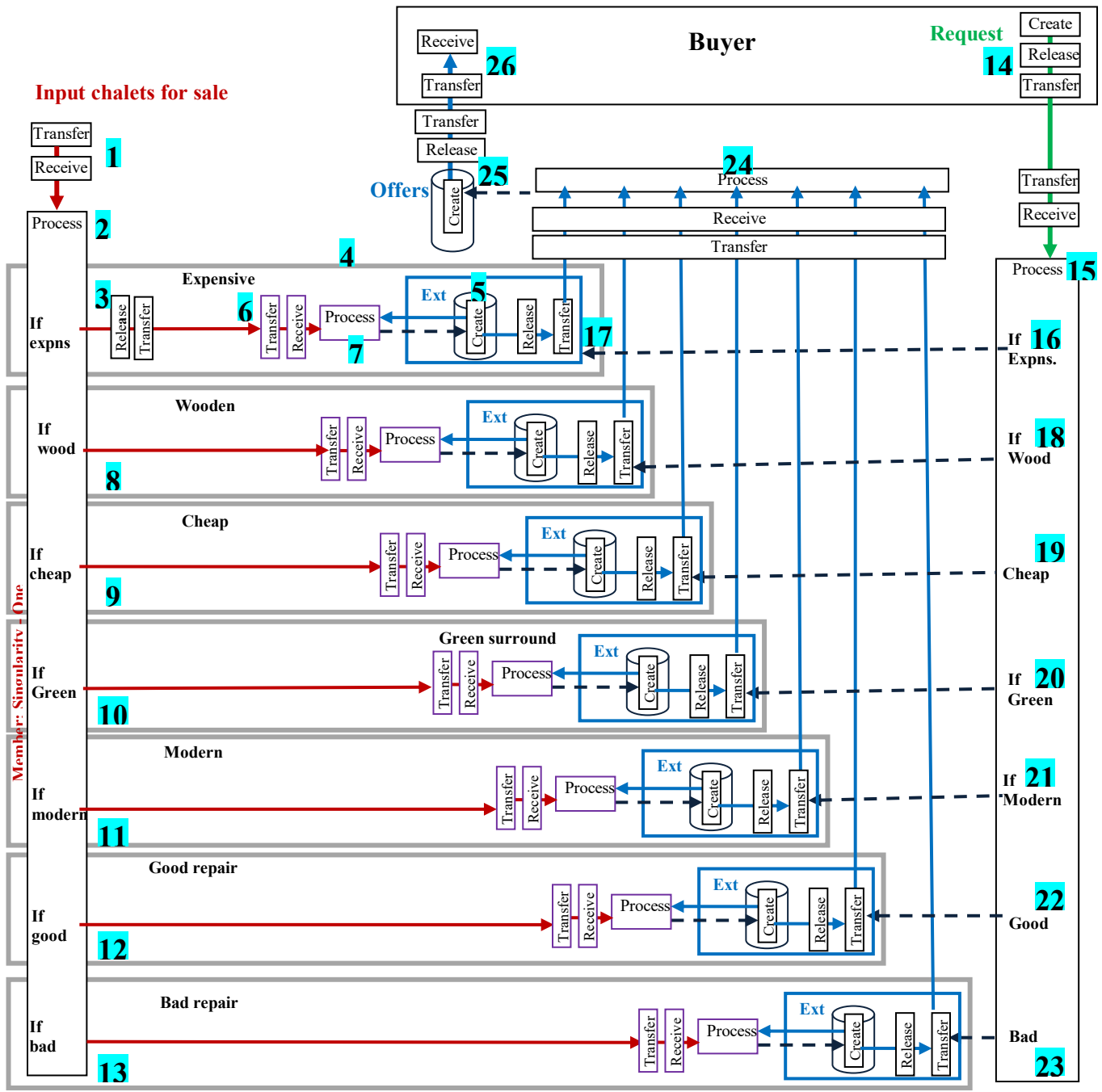

Fig. 12 The static model of the chalets market system.

A similar process is applied to the other six sets of thimacs (8-13): Each input chalet is added to its sets.

When a customer searches for a chalet to buy, they input a request (14) that is processed (15) and, according to the specification in the request, the related set is retrieved.

For example, if one of the requirements in the buyer's request is "expensive" (16), then this triggers the release (17) of the set of expensive chalets in the database.

A similar process is applied to the other six sets of thimacs (18-23): each set that satisfies one of the requirements is retrieved.

All the retrieved sets are processed (24) to create a set of chalets that satisfies all the buyer's requirements (25), set to the prospective buyer (26).

B. *Dynamic Model*

Fig. 13 shows the dynamic model. Fig. 14 shows the behavior model of the resultant chalet system.

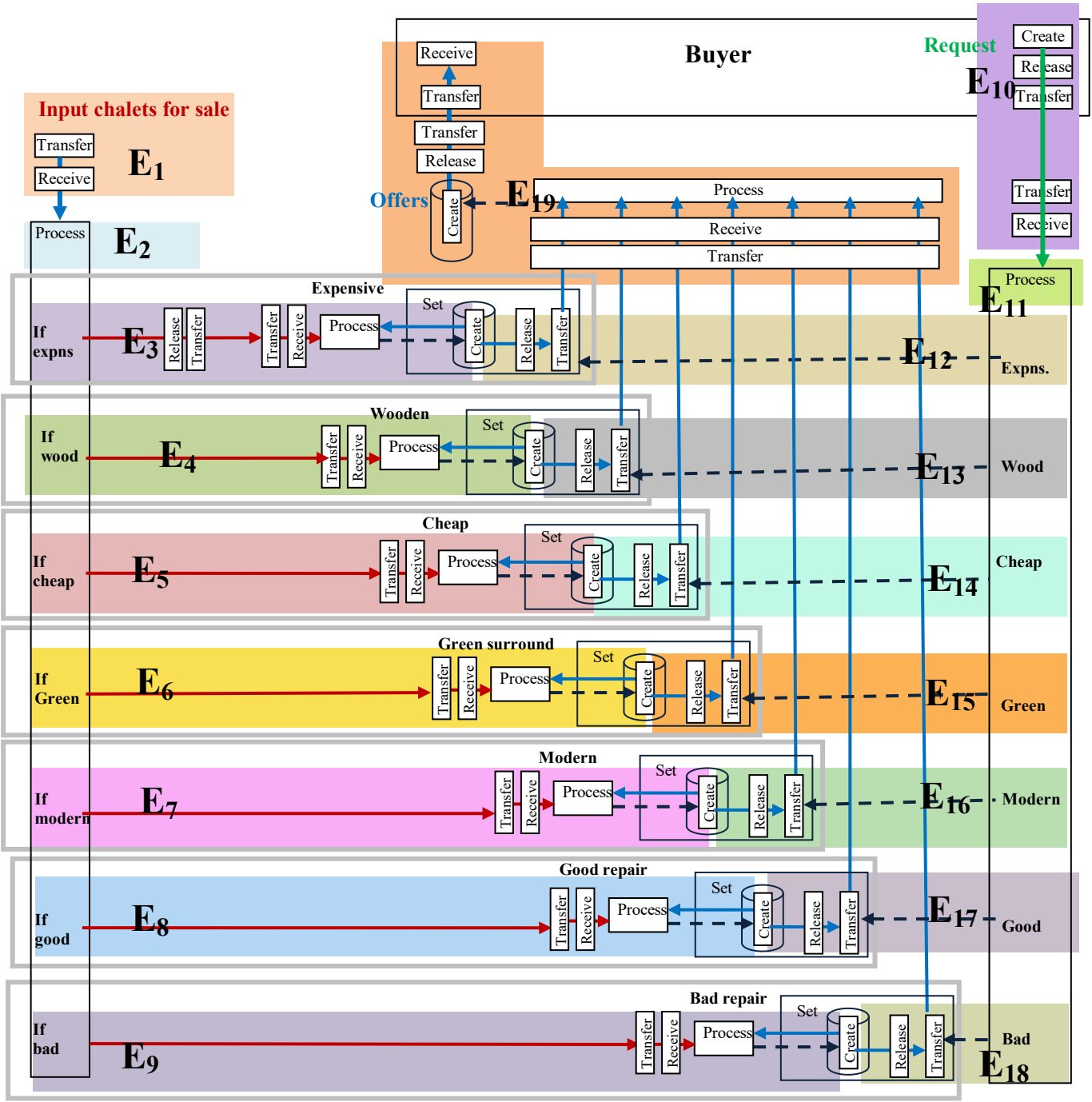

Fig. 13 The dynamic model of the chalet market system.

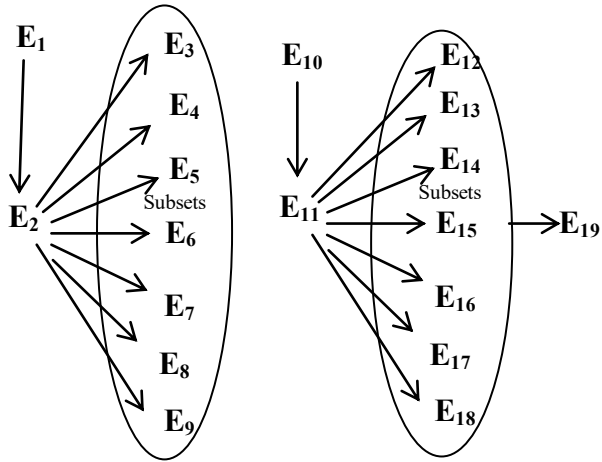

Fig. 14 The behavior model of the chalets market system.

## VII. CONCUSION

This paper is a sequel in a forward-looking roadmap of pursuing a unified conceptual modeling using TMs founded on Stoic ontology and Lupascian logic. The underlining hypothesis is that a single diagrammatic language can be applied to all types and aspects of modeling for various applications.

The roadmap consists of
- A single-category ontology called *thimac* with dual mode of being: thing and machine.
- Two levels reality of *existence* of events and *subsistence* of regions (static thimacs) (based on Stoic ontology).
- *Machines* that comprise generic actions: create, process, release, transfer and receive.
- *Things* that flow according to the structure of machines, with triggers that overrule the order of actions.
- Negative events handled based on Lupascian logic (introduced in previous publications).

The indication of this paper points to merits in the pursuit of this unifying language for the following reasons.
- The achievement of representing additional systems (e.g., businesses such as car hire systems, mathematical systems such as sets).
- The apparently successful incorporation of philosophical notions (e.g., potentiality, actuality) in TM modeling.

Accordingly, we will continue the refinement of the TM model in representing business and engineering system and expanding its philosophical foundation (e.g., ontology).